\documentclass[manuscript, nonacm]{acmart}

\usepackage{soul}
\usepackage{xspace}
\usepackage{algorithm}
\usepackage{xcolor}
\usepackage{caption}
\usepackage{booktabs}
\usepackage{multirow}
\usepackage{amsmath}
\usepackage{adjustbox} \usepackage{stfloats} \usepackage{paralist} 
\usepackage{shortcuts}

\renewcommand{\paragraph}{\vspace{3pt}\noindent\textbf}
\setcopyright{none} \acmJournal{TOPS}
\acmYear{2021}

\begin{document}

\title[\MLalgorithm Detection and Clustering of Malicious TLS Flows]{\MLalgorithm Detection and Clustering of Malicious TLS Flows}
\author{Gibran Gomez}
\email{gibran.gomez@imdea.org}
\affiliation{
  \institution{IMDEA Software Institute \& Universidad Polit{\'e}cnica de Madrid}
  \city{Madrid}
  \country{Spain}
}
\author{Platon Kotzias}
\email{platon.kotzias@gendigital.com}
\affiliation{
  \institution{Norton Research Group}
  \city{Paris}
  \country{France}
}
\author{Matteo Dell'Amico}
\email{matteodellamico@gmail.com}
\affiliation{
  \institution{University of Genoa}
  \city{Genoa}
  \country{Italy}
}
\author{Leyla Bilge}
\email{leylya.yumer@gendigital.com}
\affiliation{
  \institution{Norton Research Group}
  \city{Paris}
  \country{France}
}
\author{Juan Caballero}
\email{juan.caballero@imdea.org}
\affiliation{
  \institution{IMDEA Software Institute}
  \city{Madrid}
  \country{Spain}
}

\begin{abstract}
Malware abuses TLS to encrypt its malicious traffic, 
preventing examination by content signatures and deep packet inspection. 
Network detection of malicious TLS flows is an important,
but challenging, problem.
Prior works have proposed supervised machine learning detectors 
using TLS features.
However, by trying to represent all malicious traffic, 
supervised binary detectors produce models that are too loose,
thus introducing errors.
Furthermore, they do not distinguish flows generated by 
different malware.
On the other hand, supervised multi-class detectors 
produce tighter models and can classify flows by malware family, 
but require family labels, 
which are not available for many samples.

To address these limitations, this work proposes a novel 
\mlalgorithm approach to detect and cluster malicious TLS flows. 
Our approach takes as input network traces from sandboxes. 
It clusters similar TLS flows using \numfeatures that capture 
properties of the 
TLS client, TLS server, certificate, and encrypted payload; and 
uses the clusters to build  
an \mlalgorithm detector that can assign a malicious flow to the 
cluster it belongs to, or determine it is benign.
\updated{We evaluate our approach using 972K traces from a commercial sandbox 
and 35M TLS flows from a research network.
Our clustering shows very high precision and recall with an 
F1 score of 0.993. 
We compare our unsupervised detector with two 
state-of-the-art approaches, showing that it outperforms both.}
The false detection rate of our detector is 0.032\% measured
over four months of traffic.

\end{abstract}



\maketitle
\section{Introduction}
\label{sec:intro}

Transport Layer Security (TLS) is the most popular cryptographic protocol, 
widely used to provide confidentiality, integrity, and authentication 
to applications such as Web browsing, email, and 
instant messaging~\cite{kotzias2018coming,anderson2019tls}. 
Its security properties and wide availability 
make TLS also appealing for malware, 
which can abuse it to hide their malicious traffic among benign traffic 
of a myriad of applications, while preventing examination of its 
application payload.
In February 2020, 23\% of malware were using TLS~\cite{mal_tls_adoption}.

Network detection of malicious TLS flows is an important, 
but challenging, problem. 
It allows to protect whole networks by monitoring their traffic, 
regardless of whether endpoint security has been deployed.
An important property for scalability and privacy is that
the network detection should not require decrypting the content,
i.e., should avoid man-in-the-middle (MITM) 
interception~\cite{callegati2009mitm}. 
Another important property is to cluster the detected flows with other 
similar malicious flows, 
providing valuable threat intelligence to the analysts that 
investigate a detection.

Anderson and McGrew have proposed supervised machine learning (ML) detectors 
for malicious TLS flows~\cite{anderson2016deciphering,anderson2016identifying,anderson2017machine}. 
A limitation of binary supervised detectors 
(e.g.,~\cite{anderson2016deciphering}) 
is that they try to distinguish any malicious TLS flow, 
regardless of the malware family producing it. 
This is problematic because different families may exhibit significant 
differences in TLS usage 
(e.g., TLS versions, ciphersuites, extensions, certificates).
To cover all those differences, 
the generated model tends to become too loose, 
thus introducing errors.
In addition, a binary detector does not provide contextual information 
about similar malicious flows.
They have also proposed combining TLS features with HTTP and DNS features to 
improve the binary detection~\cite{anderson2016identifying}. 
However, some families may only use TLS 
(e.g., HTTPS, but no HTTP) and may connect using IP addresses instead of 
domain names. 
Furthermore, HTTP and DNS features may contain sensitive user information, 
thus lowering user privacy.
In their original work, Anderson et al. also evaluated a multi-class 
supervised classifier, where each class corresponds to a different malware 
family~\cite{anderson2016deciphering}. 
This approach better models the TLS traffic of individual families
and classifies the detected flows.
But, it requires clean family labels to train the classifiers.
A common labeling method is a vote on the families present in the 
AV detection labels of a sample~\cite{avclass};
unfortunately, this approach is ineffective for many samples.
 As a concrete example, Anderson and McGrew~\cite{anderson2016deciphering}
could only label this way 27\% of 20.5K samples using TLS;
no classifier could be built for the families of the 
73\% unlabeled samples, and thus their malicious traffic could not be detected. 
\updated{
Recent works have proposed anomaly based intrusion detection systems using 
neural network auto-encoders~\cite{kitsune,bovenzi2020hierarchical}.
These works do not specifically target TLS flows, 
but can be applied to them as they do
not require access to unencrypted payload.
Since they build models of benign traffic, 
they do not require labeled malicious traffic during training.
But, they suffer false positives when 
natural changes affect the benign traffic.
}

In this paper, we present a novel \mlalgorithm approach to detect and cluster
malicious TLS flows. 
Our approach respects user privacy as it only requires access to the 
encrypted TLS flows, 
and not to their unencrypted payload or any other unencrypted traffic.
Our approach takes as input traces of network traffic generated by 
executing suspicious samples in a sandbox.
From each TLS flow in the traces, it extracts \numfeatures that 
capture characteristics of the TLS client, the TLS server, 
the server's certificate, and the encrypted payload. 
After filtering benign TLS traffic, 
the remaining vectors are clustered, so that each cluster contains 
similar traffic belonging to a (potentially unknown) malware family. 
Since malware families may use different types of traffic
(e.g., C\&C, updates from download server) 
multiple clusters can be output for a family.
The use of clustering removes the requirement for family labels as
it allows detecting any malicious traffic similar to the training flows,
even if the training flows could not be annotated with a known family name,
i.e., flows from the 73\% samples that
Anderson and McGrew~\cite{anderson2016deciphering}
had to remove from their training.
The clusters are input to the \mlalgorithm detector
that can be deployed at the boundary of a network to 
identify malicious TLS flows.
The detector measures the distance of a given flow to all
clusters and outputs the closest cluster. 
If no cluster is close,
the flow is determined to be benign. 
If family labels are available for the identified cluster,
the analyst also obtains the family of the detected flow. 
When family labels are not available, the detection is not affected:
the flow is associated with the random identifier of its cluster, 
but still provides contextual information about samples generating similar 
flows.

To evaluate our approach, we use 972K network traces 
provided by a commercial sandbox vendor and 35M TLS flows
collected at the boundary of a research network over seven months.  
We identify, for the first time, how the sandbox can 
significantly impact the collected TLS traffic if it runs old OS versions 
(e.g., Windows 7, Windows XP). 
Those OSes use TLS 1.0 by default, 
instead of the currently dominating TLS 1.2 and 1.3 versions.
Thus, training a classifier with malicious traffic from a single sandbox 
using an old OS could incorrectly capture that TLS 1.0 traffic is malicious 
and TLS 1.2 and 1.3 benign.
This is an important finding because popular sandboxes 
(e.g., VirusTotal~\cite{vtOS}) 
run decade-old OS versions since a common belief is that the lack of newer OS 
defenses makes it easier for malware to run and manifest its behaviors.
Our results highlight the importance of using a variety of OS versions
in malware sandboxes.
Furthermore, this issue could have impacted 
previous work that identified significant 
differences between malware and benign programs TLS client characteristics 
(e.g., malware using older ciphersuites and less extensions)
~\cite{anderson2016deciphering,anderson2016identifying}, 
as those same authors have recently concluded that 
TLS client features are not enough by themselves~\cite{anderson2019tls}, 
in contrast with their earlier claims.

Our clustering achieves a F1 score of 0.993.
We observe that 31\% of the produced clusters only contain samples for 
which the state-of-the-art AVClass labeling tool~\cite{avclass} is not 
able to obtain family names:
supervised multi-class approaches would not work for those samples.
We also observe that our clustering is able to group TLS 1.3 flows from 
multiple samples of the same family, even when no Server Name Indication (SNI) header is present.
TLS 1.3 is a challenging case, not evaluated in prior work, 
as certificates are encrypted and 
client and sever features are greatly reduced. 

\updated{
We compare our \mlalgorithm detector with 
two state-of-the art approaches. 
Our comparison with Joy~\cite{joy}, 
the state-of-the-art supervised binary 
detector by \citet{anderson2016deciphering},
shows that when applied to the same dataset, 
\tool achieves a F1 score of 0.91, compared to 0.82 for Joy.}
\updated{We also compare \tool with Kitsune~\cite{kitsune}, 
the state-of-the-art auto-encoder-based anomaly detector.
\Tool achieves a F1 score of 0.99 compared to 0.59 for Kitsune.
}
We also evaluate our detector over long windows of time to estimate
its false detection ratio (FDR).
Over one week of traffic from the research network,
\tool achieves an FDR of 0.031\%. 
Over four months, the FDR remains almost the same at 0.032\%, 
highlighting the stability of the detection model.

This paper provides the following contributions:

\begin{itemize}

\item We present a novel \mlalgorithm approach to
detect and cluster malicious TLS flows.
Compared to binary supervised detectors,
our clustering approach models separately TLS characteristics 
from different families. 
This results in tighter models that improve detection and 
can provide contextual information about the cluster a detected flow belongs to.
Compared to multi-class classifiers, 
our approach can detect samples for which family labels are not available.

\item We observe that training malicious TLS detectors on 
traces from a single sandbox that uses old OS versions 
can significantly bias the detector.
This highlights the importance of using a variety of OS versions 
in malware sandboxes. 

\item We evaluate our approach using 972K network traces from 
a commercial sandbox and 35M TLS flows from a research network. 
Our unsupervised detector achieves a F1 score of 0.91,
compared to 0.82 for the state-of-the-art supervised detector, 
and a FDR of 0.032\% over four months of traffic.

\end{itemize}

\updated{The remainder of this paper is organized as follows. 
Section~\ref{sec:overview} motivates our research problem.
Section~\ref{sec:approach} details our 
novel unsupervised approach to detect and cluster malicious TLS flows.
Section~\ref{sec:datasets} describes the datasets used. 
Section~\ref{sec:evaluation} evaluates our approach and compares it 
with state-of-the-art approaches.
Section~\ref{sec:related} presents prior related work. 
Section~\ref{sec:discussion} discusses limitations and avenues for improvement. 
Finally, Section~\ref{sec:conclusion} concludes.
For the reader's benefit 
Table~\ref{tab:acronyms} in the Appendix details the acronyms used 
in this work.
}

\section{Motivation}
\label{sec:overview}

Our goal is to detect malicious TLS flows 
(TLS sessions)
between an infected host in a protected network, 
e.g., an enterprise or university network,
and a remote malicious server. 
A pre-requisite is that to respect user privacy only TLS flows are 
accessible from the protected network. 
This implies that the application payload of the TLS flows should not 
be accessed, i.e., no MITM,  and 
that unencrypted traffic should not be needed for the detection. 
Thus, detection features should exclusively come from TLS flows.

Our intuition is that it is possible to capture TLS characteristics 
of a specific type of malware family traffic (e.g., C\&C, updates),
but that it is very hard to capture TLS characteristics
that distinguish any malware using TLS from any benign application using TLS.
For example, certificates and domains in SNI headers are clearly 
family-specific. 
Similarly, encrypted payload features such as packet sizes are specific 
to the protocols (e.g., C\&C, update) used by each family~\cite{dietrich2013}.

Thus, rather than building a supervised binary classifier, 
we propose an \mlalgorithm detector 
that clusters similar malicious TLS flows and then detects new TLS flows by 
measuring the distance to the clusters. 
A benefit of our clustering approach is that the model
can assign detected flows to the cluster 
that led to the detection. 
Some clusters will be labeled with a recognizable family name such as 
\textit{upatre}, \textit{zbot}, or \textit{bublik}.
For others, the family may be unknown, but the cluster still provides 
important contextual information in the form of samples that generate similar 
TLS flows.
In contrast with multi-class supervised classifiers 
(e.g.,~\cite{anderson2016deciphering})
family labels are not required for the input samples,
so that the detection still works for the large fraction of samples 
that may miss them.

\paragraph{Ethical considerations.}
The passive data collection performed at the research network was 
approved and performed according to the institutional policies. 
To protect user privacy, it covers only the collection of encrypted 
TLS flows and excludes personally identifiable information  
such as client IP addresses. 
Access to the data is limited to employees of the Institution.
This research made no attempt to decrypt the TLS flows. 
Our goal is to enable the detection of malicious TLS traffic,
while maintaining user privacy.

\section{Approach}
\label{sec:approach}

\begin{figure}
  \centering
  \includegraphics[width=\columnwidth]{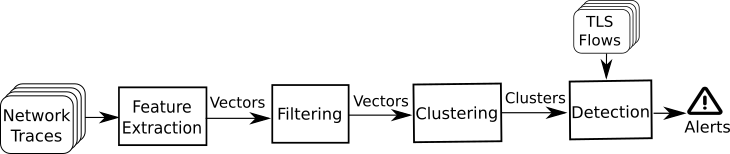}
  \caption{Approach architecture.}
  \label{fig:arch}
\end{figure}

Our approach takes as input network traces 
produced by running suspicious executables in a sandbox.
It outputs an \mlalgorithm detector 
that can be deployed on a network to detect malicious TLS flows.
The input network traces are annotated with the hash of the sample whose
execution produced it. 
Our approach comprises four steps, illustrated in Figure~\ref{fig:arch}.
Section~\ref{sec:features} describes the feature extraction that 
produces a feature vector 
for each TLS flow in the input network traces. 
Section~\ref{sec:approach-filtering} presents the filtering that 
removes feature vectors corresponding to benign traffic. 
Section~\ref{sec:clustering} details the clustering, 
which groups similar feature vectors together. 
Finally, Section~\ref{sec:detection} describes the 
\mlalgorithm detector,
which given the feature vector of a previously unseen
TLS flow classifies it as malicious (with its corresponding cluster) or benign.

\subsection{Feature Extraction}
\label{sec:features}

TLS fingerprints are applied to the first payload bytes of each 
TCP connection to identify TLS flows~\cite{dreger2006dynamic},
regardless of the ports used for the communication.
For the identified TLS flows, 
the TCP connection is reassembled,
the full TCP payload extracted, and
then the early part of the TCP payload corresponding to the TLS handshake is 
separated from the application data using 
the value of the \textit{content type} field of the TLS records.
TLS flows that have no application data in either direction are removed.

We extract \numfeatures from the remaining TLS flows. 
Of those, \numnovelfeatures are new, while the other 
\numknownfeatures have been used in prior works.
The features can be grouped into four categories.
Client, server, and certificate features are extracted from the TLS handshake,
while encrypted payload features are instead extracted from the encrypted 
application data.
Features are either numerical or categorical. To build the feature vectors,
numerical features are normalized using their z-score, i.e., by subtracting
the mean and dividing by the standard deviation.  Categorical features are
first applied one-hot encoding 
and the result is
multiplied by the term-frequency inverse document-frequency (TF-IDF) of 
the values.
\begin{table}[t]
\centering
\footnotesize
\begin{tabular}{l c l c l}
    \toprule
\textbf{Feature}  & \textbf{Type} & \textbf{PriorWork} & \textbf{TLS 1.2} & \textbf{TLS 1.3} \\
\midrule
c\_version                        & Cat & ~\cite{anderson2016deciphering}	& \checkmark & \checkmark (C) \\ c\_record\_version                & Cat & \N       & \checkmark & \checkmark \\
c\_supported\_versions            & Cat & \N       & \xmark     & \checkmark (C) \\ c\_ciphers                        & Cat & ~\cite{anderson2016deciphering}	& \checkmark & \checkmark (R) \\ c\_comp\_methods                  & Cat & \N       & \checkmark & \xmark \\ c\_curves                         & Cat & ~\cite{anderson2019tls}		& \checkmark & \checkmark (R,E) \\ c\_point\_formats                 & Cat & ~\cite{anderson2019tls}		& \checkmark & \checkmark (R,E) \\ c\_extensions                     & Cat & ~\cite{anderson2016deciphering}	& \checkmark & \checkmark (R,E) \\
c\_server\_name                   & Cat & ~\cite{malwarePrasse}			& \checkmark & \checkmark (E) \\
c\_alpn\_list                     & Cat & \N       & \checkmark & \checkmark (E) \\ c\_fake\_resumption               & Cat & \N       & \checkmark & \checkmark (C) \\ \bottomrule
\end{tabular}
\caption{Client features. (R)=Reduced. (C)=Changed. (E)=Could be Encrypted}
\label{tab:features_client}
\end{table}

\paragraph{Client features.}
These 11 features, summarized in Table~\ref{tab:features_client},
are extracted from the Client Hello message.
They capture the functionality supported by the TLS client software.
Programs either use the default 
configuration of a cryptographic library or OS API, or
configure them with their preferences.
Client features identify programs whose TLS functionality is 
configured similarly. 
The features correspond to 
the highest supported TLS and record versions,
the list of supported TLS versions (extension added in TLS 1.3),
the list of supported ciphers, compression methods, 
elliptic curves and point formats, 
the list of extensions included in the message, 
the domain name in the SNI extension, and
the list of supported application protocols in the 
Application Layer Protocol Negotiation (ALPN) extension 
(e.g., HTTP/0.9, SPDY/1).
The fake resumption feature is explained later in the resumed sessions 
paragraph.

\begin{table}[t]
\centering
\footnotesize
\begin{tabular}{l c l c l}
  \toprule
  \textbf{Feature} & \textbf{Type} & \textbf{PriorWork} & \textbf{TLS 1.2} & \textbf{TLS 1.3} \\
  \midrule
s\_dst\_port & Cat & ~\cite{anderson2016deciphering} & \checkmark & \checkmark \\
s\_version & Cat & \N & \checkmark & \checkmark (C) \\ s\_record\_version & Cat & \N & \checkmark & \checkmark \\
s\_cipher & Cat & ~\cite{anderson2016deciphering} & \checkmark & \checkmark (R) \\
s\_comp\_method & Cat & \N & \checkmark & \xmark \\ s\_extensions &  Cat & ~\cite{anderson2016deciphering} & \checkmark & \checkmark (R,E) \\ s\_alpn\_list & Cat & ~\cite{anderson2019tls} & \checkmark & \checkmark (E) \\ s\_session\_ticket\_lifetime & Cat & ~\cite{anderson2019tls} & \checkmark & \checkmark \\
s\_ct\_signature & Cat & \N & \checkmark & \checkmark (E) \\ \bottomrule
\end{tabular}
\caption{Server features.}\label{tab:features_server}
\end{table}

\paragraph{Server features.}
The 9 server features in Table~\ref{tab:features_server} 
correspond to the destination port and 
features extracted from the Server Hello message.
Server features capture the TLS functionality of the server software, 
i.e., the parameters the server selects for the TLS session 
after intersecting the client TLS support 
with its own TLS support.
Server features identify servers configured similarly.
The Server Hello features are the selected 
TLS version, record versions, cipher, and compression method; 
the list of extensions in the message,
the selected application protocol in the ALPN extension,
and the lifetime in the SessionTicket extension.
The last feature captures the signature in the 
SignedCertificateTimestamp extension that a server may use to transmit 
signed proofs of the server's certificate presence in 
the Certificate Transparency (CT) logs~\cite{rfc6962}.

\begin{table}[t]
\centering
\footnotesize
\begin{tabular}{l c l c c}
  \toprule
  \textbf{Feature} & \textbf{Type} & \textbf{PriorWork} & \textbf{TLS 1.2} & \textbf{TLS 1.3} \\
  \midrule
(c$|$s)\_num\_certs & Cat & ~\cite{anderson2016deciphering} & \checkmark & \xmark \\
(c$|$s)\_leaf\_cert\_version & Cat & \N & \checkmark & \xmark \\
(c$|$s)\_leaf\_cert\_validity & Cat & ~\cite{anderson2016deciphering} & \checkmark & \xmark \\ (c$|$s)\_leaf\_cert\_num\_SAN & Cat & ~\cite{anderson2016deciphering} & \checkmark & \xmark \\
(c$|$s)\_leaf\_cert\_ext\_num & Cat & \N & \checkmark & \xmark \\
s\_leaf\_cert\_validation\_status & Cat & \N & \checkmark & \xmark \\ s\_leaf\_cert\_self\_signed & Cat & ~\cite{anderson2016deciphering} & \checkmark & \xmark \\ (c$|$s)\_leaf\_cert\_sign\_alg & Cat & ~\cite{anderson2016deciphering} & \checkmark & \xmark \\
(c$|$s)\_leaf\_cert\_pubkey\_hash & Cat & \N & \checkmark & \xmark \\
(c$|$s)\_leaf\_cert\_pubkey\_size & Cat & ~\cite{anderson2016deciphering} & \checkmark & \xmark \\
(c$|$s)\_leaf\_cert\_subj\_cn & Cat & \N & \checkmark & \xmark \\
(c$|$s)\_leaf\_cert\_subj\_o & Cat & \N & \checkmark & \xmark \\
(c$|$s)\_leaf\_cert\_subj\_ou & Cat & \N & \checkmark & \xmark \\
(c$|$s)\_leaf\_cert\_subj\_c & Cat & \N & \checkmark & \xmark \\
(c$|$s)\_leaf\_cert\_subj\_st & Cat & \N & \checkmark & \xmark \\
(c$|$s)\_leaf\_cert\_subj\_l & Cat & \N & \checkmark & \xmark \\
(c$|$s)\_leaf\_cert\_subj\_email & Cat & \N & \checkmark & \xmark \\
(c$|$s)\_leaf\_cert\_iss\_cn & Cat & \N & \checkmark & \xmark \\
(c$|$s)\_leaf\_cert\_iss\_o & Cat & \N & \checkmark & \xmark \\
(c$|$s)\_leaf\_cert\_iss\_ou & Cat & \N & \checkmark & \xmark \\
(c$|$s)\_leaf\_cert\_iss\_c & Cat & \N & \checkmark & \xmark \\
(c$|$s)\_leaf\_cert\_iss\_st & Cat & \N & \checkmark & \xmark \\
(c$|$s)\_leaf\_cert\_iss\_l & Cat & \N & \checkmark & \xmark \\
(c$|$s)\_leaf\_cert\_iss\_email & Cat & \N & \checkmark & \xmark \\
\bottomrule
\end{tabular}
\caption{Certificate features.}
\label{tab:features_cert}
\end{table}

\paragraph{Certificate features.}
These 24 features, summarized in Table~\ref{tab:features_cert}, 
are extracted from the certificate chain sent by the server.
These features capture the number of certificates in the chain and
fields of the leaf certificate.
The focus is on the leaf certificate because that certificate is specific to the
service, while other certificates in the chain belong to the 
certification authorities (CAs) used, and thus
may be common to many unrelated services. 
Nine leaf certificate features correspond to certificate parameters,
namely 
the version, 
its validity period, 
the number of Subject Alternative Names (SAN) and extensions included, 
the validation status on the day we first process a flow, 
whether the certificate is self-signed, 
the signature algorithm, 
public key length, and 
public key hash. 
Another seven features correspond to fields of the Subject 
Distinguished Name (DN), and the remaining seven to the same fields 
in the Issuer DN.
In the rare case where client certificates are used 
an extra 22 analogous features are extracted from 
the client's certificate chain. 

\begin{table}[t]
\centering
\footnotesize
\begin{tabular}{l c l c c}
  \toprule
\textbf{Feature}  & \textbf{Type} & \textbf{PriorWork} & \textbf{TLS 1.2} & \textbf{TLS 1.3} \\
\midrule
enc\_data\_size & Num & \N &  \checkmark & \checkmark \\
enc\_sent\_size & Num & \N &  \checkmark & \checkmark \\
enc\_recv\_size & Num & \N &  \checkmark & \checkmark \\
enc\_num\_pkts & Num & \N &  \checkmark & \checkmark \\
enc\_sent\_pkts & Num & \N &  \checkmark & \checkmark \\
enc\_recv\_pkts & Num & \N &  \checkmark & \checkmark \\
c\_max\_seq & Num & \N &  \checkmark & \checkmark \\
c\_max\_length & Num & \N &  \checkmark & \checkmark \\
s\_max\_seq & Num & \N &  \checkmark & \checkmark \\
s\_max\_length & Num & \N &  \checkmark & \checkmark \\
sent\_recv\_pkts\_ratio & Num & \N &  \checkmark & \checkmark \\
sent\_recv\_size\_ratio & Num & \N &  \checkmark & \checkmark \\
msg\_pkts\_c\_0 & Num & \N &  \checkmark & \checkmark \\
msg\_size\_c\_0 & Num & ~\cite{dietrich2013} &  \checkmark & \checkmark \\
msg\_pkts\_s\_0 & Num & \N &  \checkmark & \checkmark \\
msg\_size\_s\_0 & Num & ~\cite{dietrich2013} &  \checkmark & \checkmark \\
msg\_pkts\_c\_1 & Num & \N &  \checkmark & \checkmark \\
msg\_size\_c\_1 & Num & ~\cite{dietrich2013} &  \checkmark & \checkmark \\
msg\_pkts\_s\_1 & Num & \N &  \checkmark & \checkmark \\
msg\_size\_s\_1 & Num & ~\cite{dietrich2013} &  \checkmark & \checkmark \\
msg\_pkts\_c\_2 & Num & \N &  \checkmark & \checkmark \\
msg\_size\_c\_2 & Num & ~\cite{dietrich2013} &  \checkmark & \checkmark \\
msg\_pkts\_s\_2 & Num & \N &  \checkmark & \checkmark \\
msg\_size\_s\_2 & Num & ~\cite{dietrich2013} &  \checkmark & \checkmark \\
\bottomrule
\end{tabular}
\caption{Payload features.}
\label{tab:features_payload}
\end{table}

\paragraph{Encrypted payload features.}
Another 24 features are extracted from the encrypted application data 
transferred after the TLS handshake has completed.
These features capture the application protocol used for communication. 
The intuition is that the protocol changes infrequently
because protocol updates require 
both client and server software updates and need to be 
thoroughly debugged to maintain compatibility. 
In particular, prior work has shown that the command and control (C\&C) 
protocol used by malware changes much slower than its communication 
endpoints (i.e., domains, IP addresses, ports)~\cite{polyglot}. 
Similarly, we expect the protocol to also change less frequently than 
the TLS configuration parameters.

Traffic analysis approaches often leverage the number, 
direction (i.e., client-to-server or server-to-client) and 
sizes of packets as features to identify 
encrypted content (e.g.,~\cite{sun2002statistical,chen2010side}).
However, packets do not always accurately capture the underlying protocol 
because application messages can be fragmented into multiple packets 
by the transport and IP protocols.
To address this issue, we define a \textit{sequence} as 
all consecutive packets sent in one direction until 
another packet is seen in the opposite direction. 
The concatenation of the payload of all packets in a sequence is a 
good approximation of an application message~\cite{dietrich2013}.
The use of variable-length sequences avoids the need to select a threshold. 

We call two consecutive sequences in opposite direction
a \textit{request-response pair} (RRP).
The number of RRPs to consider is a hyper-parameter of the payload features. 
The intuition to select this parameter is that the initial part of the 
communication is more commonly related to the protocol, 
while later parts may be more related to the transferred content 
(e.g., files sent).
In this work, we consider the first three RRPs, 
analyzing a total of six sequences. 
In our sandbox traces we observe that 95\% of TLS flows have a single RRP, 
3\% have two, and 2\% have at least three RRPs. 

For each sequence, two features are extracted: 
the size of the concatenated payload (e.g., msg\_size\_c\_0) and 
the number of packets in the sequence (e.g., msg\_pkts\_c\_0).
These sequence features correspond to half of the 24 payload features.
The other payload features correspond to the 
total byte size of the encrypted payload 
(and its split in sent and received bytes); 
the total number of packets 
(and its split in sent and received packets); 
the size of the larger sequence in each direction;
the maximum number of packets in a sequence in each direction; 
the ratio of packets sent over packets received; and 
the ratio of bytes sent over bytes received.

In conclusion, payload features identify TLS flows with 
similar content, despite domain and IP address polymorphism. 
We have also experimented with timing features, 
but have found them too sensitive to the network setup
(e.g., server location, congestion) and thus we do not use them.

\paragraph{Resumed sessions.}
TLS resumption allows to quickly re-establish a prior TLS session 
using a shorter handshake~\cite{rfc5077}.
The server encapsulates the session state into a ticket 
sent to the client. 
Later, the client can resume the previous session by sending
the corresponding ticket to the server. 
The shorter handshake does not include client or server certificates. 
To avoid leaving the certificate features empty
(which may make resumed sessions look alike), 
feature extraction tracks session tickets sent from servers.
When a TLS session is resumed, it uses the ticket sent by the 
client to identify the original TLS session 
(containing the same ticket in the opposite direction) 
and extracts the missing certificate features from the original session.
If the original session cannot be identified, 
the fake resumption client feature is set to indicate that it may not be a real resumed session.
Fake resumptions are used to avoid confusing middleboxes that do not
support TLS 1.3~\cite{fake_resumption}.
We also observe them used by \textit{Ultrasurf}, 
an Internet censorship circumvention tool,
which establishes TLS 1.2 flows without certificates.

\paragraph{TLS 1.3.}
Most features presented so far are available up to TLS 1.2. 
However, TLS 1.3 changes the protocol to reduce the 
information available in the TLS handshake. 
In particular, some fields become encrypted and other fields have 
fewer values to choose from.
These changes are captured in the TLS 1.3 columns in 
Tables~\ref{tab:features_client}--\ref{tab:features_payload}.
In particular, client features such as the list of ciphers, elliptic curves, 
and point formats provide less information in TLS 1.3, 
as many values have been removed for increased protection,
e.g., against downgrade attacks.
However, as long as clients still support TLS 1.2, we expect
the removal of those values to happen slowly over time.
Furthermore, certificates are encrypted so their features cannot be 
extracted directly from the network trace.
On the other hand, payload features are not affected.
Despite these changes, our approach is able to produce 
accurate TLS 1.3 clusters.

\subsection{Filtering}
\label{sec:approach-filtering}

Sandbox traces may include benign traffic from different sources. 
One source is background traffic generated by the OS and 
other benign programs installed in the virtual machine. 
Another source are benign samples executed in the sandbox. 
Yet another source are flows to benign services performed by 
malicious samples such as those to test Internet connectivity. 
Some of that benign traffic may use TLS. 
To identify benign traffic in the network traces we use 
the Tranco list of the top 1M popular domains~\cite{LePochat2019,trancoList}. 
Removing benign TLS flows is important to avoid generating 
clusters of benign traffic 
and for scalability.
We also filter vanilla Tor traffic used by some malware samples, 
which we identify using a previously proposed fingerprint~\cite{torbridges}.
Note that for better scalability, 
flows without application data in either direction
were already removed prior to feature extraction.
Of course, any filtering can be incomplete ,
e.g., some unpopular benign domains could remain. 
An advantage of an unsupervised model is that any
remaining benign traffic would produce its own cluster. 
Once that benign traffic is identified, the cluster can be removed, 
without requiring a retraining of the whole model, 
which would be needed by supervised approaches.

\subsection{Clustering}
\label{sec:clustering}

The goal of the clustering is to group together similar feature vectors that 
correspond to the same type of malicious TLS traffic.
Each cluster comprises of feature vectors generated by the same or 
different samples. 
Feature vectors from the same sample may end up in different 
clusters if the sample produces different types of communication such as 
C\&C communication or communication with an update server.
When a cluster contains TLS flows from different samples, 
those samples should belong to the same malware family. 
However, samples from the same family could end up in different clusters, 
e.g., when a subset of samples of the family exhibit some type of traffic 
and a different subset of family samples exhibit different traffic.

We use a hierarchical density-based clustering
algorithm based on
HDBSCAN~\cite{campello_density-based_2013}. It does not require the number of clusters be specified, 
recognizes clusters of arbitrary shape and variable
density, scales to large datasets, and 
allows working with any kind of data by defining arbitrary distance functions.
It distinguishes between clusters
and noise, i.e., scattered data points that shouldn't be considered part of any cluster.
It has three hyper-parameters: the number of elements
that should be close to a central one to define a dense zone
(\textit{mpts}), the minimum cluster size (\textit{mcs}), and a
parameter (\textit{m}) that tunes the density of linkage in the
data structure it uses for neighbor search.
Our evaluation searches for the best values for these parameters.
The distance function used divides the features in two sets: numerical
and categorical.  First, it calculates the Euclidean distance of the
numerical features, and multiplies it by the fraction of numerical
over all features.  Then, it computes the cosine distance between the
categorical features, multiplying it by the fraction of categorical
features. The final distance is the sum of these two values.

For each produced cluster, the domains in the SNI header and the
leaf certificates of the flows in the cluster are collected, 
so that they can be added to blacklists. 
In addition, our approach tries to assign a human-interpretable label 
to each cluster by applying the AVClass~\cite{avclass} labeling tool to 
the samples that produced the flows in the cluster. 
We detail the labeling in Section~\ref{sec:classification}. 
The cluster label corresponds to the family that has a majority in the cluster
followed by a suffix to differentiate multiple clusters from the same family. 
For 31\% of the clusters AVClass cannot obtain a family for any sample, 
but, in contrast to multi-class supervised classifiers, 
this does not affect our \mlalgorithm detector, 
affecting only the availability of human readable 
family names to identify the clusters.

\subsection{Detection}
\label{sec:detection}

The \mlalgorithm detector leverages the produced clustering model to
decide whether a previously unseen flow belongs to a known cluster. 
Detection consists in searching for the closest node to a given flow,
if any. Otherwise the flow is considered an outlier (i.e. benign). 
For this, we evaluate two different methods. 
The \textit{variable threshold} method determines the density of a cluster 
as the largest distance of a node in that cluster from its closest neighbor; 
if the distance of the vector to a node in a cluster is below this threshold,
the vector is considered to belong to that cluster. 
The \textit{fixed threshold} method instead defines a fixed threshold 
for all clusters, so that the unseen
element should be close enough to a cluster's node to be part of it.
Regardless of the method, if the given flow is assigned to a cluster, 
then it is labeled as malicious and the cluster identifier is output.
If no cluster is close enough, the flow is labeled as benign.
\updated{The detection threshold is the main parameter that controls 
false positives (FPs) and false negatives (FNs). 
If set very tight, then false positives are minimized, 
at the expense of increasing false negatives. 
When relaxed, false positives increase, while false negatives reduce.
We evaluate this effect in Section~\ref{sec:eval:detection}.
}

\begin{table*}[t]
\footnotesize
\begin{adjustbox}{max width=\linewidth}
\begin{tabular}{lllr rrr rrrr rrrr}
  \toprule
    
\multirow{2}{*}{\bf Dataset} & \multirow{2}{*}{\bf Start} & \multirow{2}{*}{\bf End} & \multirow{2}{*}{\bf Samples}
& \multicolumn{3}{c}{\textbf{TLS Destinations}} & \multicolumn{4}{c}{\textbf{TLS Flows}} & \multicolumn{4}{c}{\textbf{TLS Version}} \\

& & & &
\textbf{SNI} & \textbf{IP} & \textbf{Ports} &
\textbf{All} & \textbf{Payload} & \textbf{Filtered} & \textbf{Resumed} & 
\textbf{1.0} & \textbf{1.2} & \textbf{1.3} & \textbf{Other} \\
\cmidrule(r){1-4} \cmidrule(lr){5-7} \cmidrule(lr){8-11} \cmidrule(l){12-15}
	SB-small  & 2017-11-01 & 2018-01-30 & 342.1K & 49.8K & 17.9K & 268 & 1.9M & 70.1\% & 8.8\% & 19.7\% & 87.5\% & 12.3\% & 0\% & 0.2\% \\
	SB-medium  & 2019-05-30  & 2019-10-30 &  630.5K   & 65.8K  & 22.4K & 339 & 11.0M & 16.6\%  & 8.1\% & 15.0\%   & 95.4\%     & 2.9\%      & 1.7\% & 0.0\% \\
	SB-all     & 2017-11-01  & 2019-10-30 &  972.6K & 113.5K & 37.6K & 527 & 12.9M & 24.5\%  & 8.2\%  & 15.7\% & 93.5\% & 5.6\% & 0.9\% & 0.0\% \\
\bottomrule
\end{tabular}
\end{adjustbox}
\caption{Summary of network traces collected from a commercial sandbox. }
\label{tab:datasets_malware}
\end{table*}

\begin{table*}[t]
\centering
\scalebox{0.9}{
\footnotesize
\begin{tabular}{lllr rrr rrrr rrrr}
  \toprule
    
\multirow{2}{*}{\bf Dataset} & \multirow{2}{*}{\bf Start} & \multirow{2}{*}{\bf End} & \multirow{2}{*}{\bf SRC}
& \multicolumn{3}{c}{\textbf{TLS Destinations}} & \multicolumn{4}{c}{\textbf{TLS Flows}} & \multicolumn{4}{c}{\textbf{TLS Version}} \\

& & & &
\textbf{SNI} & \textbf{IP} & \textbf{Ports} &
\textbf{All} & \textbf{Payload} & \textbf{Filtered} & \textbf{Resumed} & 
\textbf{1.0} & \textbf{1.2} & \textbf{1.3} & \textbf{Other} \\
\cmidrule(r){1-4} \cmidrule(lr){5-7} \cmidrule(lr){8-11} \cmidrule(l){12-15}

Benign01  & 2019-10-08 & 2019-11-24 & 1.0K & 108.6K & 85.5K  & 824  & 9.6M  & 96.0\% & 8.9\% & 37.3\% & 0.9\% & 86.8\% & 12.2\% & 0.1\% \\
Benign02  & 2019-12-10 & 2020-01-31 & 1.0K & 127.2K & 100.3K & 1.1K & 13.2M & 96.2\% & 9.8\% & 42.5\% & 0.8\% & 80.7\% & 18.4\% & 0.1\% \\
Benign03  & 2020-02-01 & 2020-04-30 & 1,0K & 120.8K & 98.4K  & 2.9K & 11.6M & 96.0\% & 7.9\% & 43.4\% & 0.9\% & 81.7\% & 17.4\% & 0.0\% \\
Comp.Train& 2020-10-23 & 2020-10-25 &  845 &  6,257 &  8,409 &   37 & 879.9k & 66.5\% &27.9\% & 34.9\% & 0.7\% & 65.5\% & 33.7\% & 0.1\% \\
Comp.Test & 2020-10-28 & 2020-10-28 &  496 &  7,277 &  9,603 & 1.0K & 396.2k & 74.9\% &18.7\% & 36.9\% & 0.3\% & 69.0\% & 30.3\% & 0.4\% \\

\bottomrule
\end{tabular}
}
\caption{Summary of collected research network traffic.}
\label{tab:datasets_benign}
\end{table*}

\section{Datasets}
\label{sec:datasets}

To perform the clustering and build the detectors, 
we use 972K sandbox network traces provided to us by a \vendor,
summarized in Table~\ref{tab:datasets_malware}. 
To evaluate the produced detector, 
we use seven months of TLS traffic collected at the boundary of a 
research network, summarized in Table~\ref{tab:datasets_benign}. 

\paragraph{Sandbox network traces.}
We use two datasets (SB-small, SB-medium) of network traces obtained 
by a \vendor from the execution of suspicious samples in their sandbox. 
Each sample was run for one minute and sleeps introduced by the sample 
were skipped.
The network trace contains all the traffic produced by the sample.
Each sample has a single network trace.
All traces contain TLS flows because that was the selection criteria 
used by the \vendor to share executions with us.

The union of the SB-small and SB-medium datasets produces the SB-all dataset.
SB-all contains before filtering 12.9M flows on 527 destination ports, 
with the top three being
443/tcp (93.6\%), 9001/tcp (2.67\%), and 80/tcp (1.87\%).  
Of the 12.9M TLS flows, 24.5\% have a payload and 8.2\% (1M) 
remain after filtering and are used in the clustering. 
We explain the reasons for this significant drop in Section~\ref{sec:versions}.
Those 1M TLS flows are almost exclusively for 443/tcp (99.8\%) and are
originated by 
28\% of the 972K samples. 
Of all samples, 
9\% exclusively communicate through TLS (excluding DNS and DHCP)
while 91\% use HTTP in addition to TLS.
Thus, HTTP traffic could not be used for detecting 9\% of the samples, 
even if user privacy was not a concern.
We also observe three times more effective second-level domains (e2LDs) contacted through TLS compared to HTTP.
6\% of the e2LDs are contacted both via HTTP and HTTPS, 
likely due to HTTP redirections towards HTTPS URIs.

\paragraph{Research network traffic.}
To evaluate false positives (FPs), 
we use logs of TLS flows collected at the boundary of a research network.
Since no infections were detected in this network during the monitoring 
period we assume that the traffic is benign.
For privacy reasons, the logs consist exclusively of TLS flows.
The traffic is split into five datasets, 
summarized in Table~\ref{tab:datasets_benign}. 
The first three comprise over seven months of traffic originated by 
1,216 source IP addresses that produce 34M TLS flows.
For the comparison with Joy~\cite{joy}, 
we collected two additional short captures, 
where we run Joy in parallel with our TLS log collection.

\subsection{TLS Versions}
\label{sec:versions}
There exist two significant differences between the SB-all sandbox traffic 
and the benign traffic from the research network. 
First, in SB-all only 24\% of the TLS flows exchanged 
any application data.
This is in stark contrast with 96\% of the research network flows having 
application data. 
Second, 93\% of the sandbox flows after filtering use TLS 1.0, 
5.6\% TLS 1.2 and less than 1\% use TLS 1.3. 
This TLS 1.0 dominance is in stark contrast with the traffic from the 
research network where 80-86\% of the flows after filtering use TLS 1.2, 
12\%-18\% TLS 1.3, and less than 1\% TLS 1.0.
These differences cannot be due to the different dataset time frames 
as SB-medium and Benign01 partially overlap in the second half of 2019.

We believe both differences are rooted in the sandbox 
using almost exclusively Windows 7 virtual machines, 
which we have verified by applying the p0f passive fingerprinting tool on the 
network traces~\cite{p0f}. 
Windows 7 system libraries by default use TLS 1.0.
TLS 1.1 and 1.2 are supported, 
but have to be manually enabled by the user~\cite{win7}. 
We believe the two differences are caused by the majority of the malware using 
the default Windows TLS functionality, 
rather than a cryptographic library statically linked in the executable. 
If the malware was using a statically linked cryptographic library, then, 
in 2017-2018, the default would be to use TLS 1.2. 
To use TLS 1.0, the malware developers would have to configure the used 
cryptographic library to specifically use TLS 1.0, which seems unlikely 
as this happens for samples from many families.
The 6.5\% of flows using TLS 1.2 and 1.3 likely correspond to 
malware using a statically linked TLS library or invoking the 
default browser in the sandbox to open a webpage. 

The low fraction of TLS flows with application data is also likely due to the 
 use of the default TLS functionality in Windows 7. 
The malware tries to connect to
the servers using TLS 1.0, but many servers no longer support TLS 1.0, or
the offered ciphersuites, and reject it.  Note that the PCI Council
suggested that organizations migrate from TLS 1.0 to TLS 1.1 or higher before
June 30, 2018~\cite{pciDeprecation}.  This conclusion is supported by 37\% of
the TLS flows having a protocol\_version alert, i.e., no TLS version can be
agreed, and another 16\% a handshake\_failure alert, i.e., no
ciphersuite can be agreed.

This analysis highlights the impact sandbox configuration can have on the 
collected data. 
Configuring the sandbox with old software may be beneficial to observe some 
behaviors like exploitation, but it can negatively impact other aspects 
such as TLS behavior.
Running the sample on multiple OSes is arguably the best alternative, 
but it impacts scalability.
Note that we do not control the sandbox and thus cannot 
configure it with newer OSes. 
This is a common scenario where data collection and analysis are performed 
by different teams.
The main impact of the sandbox in our work is that flows with no 
application data are filtered, and eventually some network traces 
may be removed because they do not contain any flows with 
application data. 
Thanks to the large size of our dataset, even after filtering, 
we are still left with 1M TLS flows from 272K malicious samples.
We do not think other selection biases are introduced as the TLS version 
has little influence in the clustering results according 
to the feature analysis process.

\paragraph{TLS 1.3.}
Since its release on August 2018 TLS 1.3 has seen fast 
adoption~\cite{kotzias2018coming,holz2019era}.
Holtz et al. measured that by April 2019, 
it was used in 4.6\% of TLS flows~\cite{holz2019era}.
In our research network, one year later we observe 17.4\% TLS 1.3 flows,
nearly a fourfold increase.
In the sandbox dataset, after filtering a modest 0.9\% flows from 6.8K 
samples use TLS 1.3. 
This shows that a minority of malware authors try to keep their TLS traffic 
as secure as possible, avoiding the default TLS versions offered by the OS and 
using instead a statically linked cryptographic library they can customize.
Most TLS 1.3 flows belong to three malware families: 
\textit{sofacy}, 
an alias for the Fancy Bear (APT28) Russian cyber espionage group,
\textit{lockyc}, an imitator of the \textit{locky} ransomware, and
\textit{razy} that steals cryptocurrency wallets~\cite{razy}.
The main impact of TLS 1.3 is that certificates are not observable.
Additionally, the SNI can be transmitted encrypted by using new extensions.
We also expected a reduced number of ciphersuites, 
but due to backwards compatibility we observe 14--78 ciphersuites being offered,
much higher than the five standard TLS 1.3 ciphersuites.
Despite the reduced feature set, 
our evaluation shows our approach successfully handles TLS 1.3.

\subsection{Family Labeling}
\label{sec:classification}

We query the hashes of all 972K sandbox samples to VirusTotal (VT)
to collect their AV labels.
Among those, 64\% (623K) were known to VT. We feed the VT reports to the AVClass malware labeling tool~\cite{avclass}.
AVClass outputs the most likely family name for each sample and also
classifies it as malware or PUP based on the presence of PUP-related keywords
in the AV labels (e.g., \textit{adware}, \textit{unwanted}).
Overall, AVClass labels 59\% (574K) of the samples.
For the remaining samples no family was identified because their labels 
were generic.
The 272K samples in the filtered SB-all dataset are grouped in 
738 families, 545 malware and 193 PUP.
For the interested reader, Table~\ref{tab:top_fams} in the Appendix 
shows the top 20 families.

To establish the malware family responsible for a cluster, 
we apply a majority vote on the AVClass labels of the samples in the cluster.
However, for 35\% of clusters, AVClass does not assign a family to 
any of the samples in the cluster 
(i.e., they only have generic labels), and 
thus the cluster cannot be labeled.
Note that, 
in contrast to multi-class supervised classifiers, 
our \mlalgorithm detector
does not use family labels and thus
can still detect flows for the 41\% of samples and 35\% of clusters 
that AVClass cannot label. 
In a detection, even when the cluster to which the detected flow belongs 
does not have an associated family name, 
it still provides important contextual information to the analysts
by capturing other malicious samples that generate similar traffic.

\begin{table}[t]
\centering
\footnotesize
\begin{tabular}{l r r r}
\toprule
\textbf{Family} & \textbf{Total flows} & \textbf{Samples} & \textbf{Clusters} \\
\midrule
clipbanker  & 10,183 &  9,718 &  3 \\
shiz        & 10,105 &  9,250 &  4 \\
upatre      & 20,352 &  8,738 &  3 \\
bublik      &    140 &     57 &  5 \\
---         &    185 &    170 & 12 \\
ekstak      &     20 &     20 &  1 \\
miancha     &      9 &      9 &  1 \\
\midrule
Total   & 40,994 & 27,962 & 29 \\
\bottomrule
\end{tabular}
\caption{\small{Manually generated ground truth.}}
\label{table:groundtruth_clusters}
\end{table}

\subsection{Ground Truth}
\label{sec:gt}

To evaluate the clustering results,
we use a manually labeled subset of flows from SB-all, 
summarized in Table~\ref{table:groundtruth_clusters}. 
AVClass results were used to randomly select 
samples from four of the largest families
(\textit{clipbanker}, \textit{upatre}, \textit{shiz}, \textit{bublik}),
a couple of small families (\textit{ekstak}, \textit{miancha}), and
a variety of samples that AVClass cannot classify.
Note that the AVClass family is not used as ground truth by itself, 
as it is not available for all samples, 
it may be incorrect for some samples, and does not separate 
different types of traffic from the same sample.
Still, the family labels help ensure a variety of malware is included.
Then, the benign traffic was filtered and features were extracted 
(e.g., certificates, domains in SNI headers, payload sequences).
In addition, additional information was obtained that 
is not available to our detector such as 
VT reports that contain file properties and behavioral information,
other features in the network traces not used by our approach
(e.g., non-TLS traffic, destination IP addresses), and the
AVClass family.
By examining static, dynamic, and shared indicators in the reports and the 
network traffic, 29 clusters were identified.

\section{Evaluation}
\label{sec:evaluation}

This section first presents the clustering results 
in Section~\ref{sec:eval:clustering} and 
then the detection results in Section~\ref{sec:eval:detection}.

\begin{table*}[t]
\centering
\footnotesize
\begin{tabular}{llll rr rrrrr rrr r}
  \toprule
  \multicolumn{4}{c}{\bf Clustering} & \multicolumn{2}{c}{\bf Clusters} & \multicolumn{5}{c}{\bf Cluster Size} & \multicolumn{3}{c}{\bf Accuracy} & \multirow{2}{*}{\bf Time} \\
\textbf{ID} & \textbf{Features} & \textbf{Param.} & \textbf{ZS} & \textbf{All} & \textbf{Singl.} & \textbf{Min} & \textbf{Max} & \textbf{Med.} & \textbf{Mean} & \textbf{PStdev} & \textbf{Prec.} & \textbf{Recall} & \textbf{F1} & \\

\cmidrule(r){1-4} \cmidrule(lr){5-6} \cmidrule(lr){7-11} \cmidrule(lr){12-14} \cmidrule(l){15-15}

FD1 & all & default & \Y & 27 & 0 & 2 & 9,978 & 65 & 1,518.3 & 3,121.6 & 0.993 & 0.872 & 0.928 & 18.0 \\
FD2 & no-client & default & \Y  & 25 & 0 & 2 & 9,990 & 65 & 1,639.8 & 3,257.3 & 0.992 & 0.877 & 0.931 & 16.9 \\
FD3 & no-server & default & \Y & 10 & 0 & 21 & 20,213 & 354 & 4,099.4 & 6,274.6 & 0.747 & 0.905 & 0.819 & 16.6 \\
FD4 & no-cert & default & \Y & 27 & 0 & 2 & 10,003 & 26 & 1,518.3 & 3,503.6 & 0.996 & 0.990 & 0.993 & 13.6 \\
FD5 & no-payload & default & \Y & 9 & 0 & 51 & 10,290 & 4,951 & 4,554.9 & 4,196.0 & 0.982 & 0.855 & 0.914 & 13.1 \\
FD6 & payload & default & \Y & 20 & 0 & 4 & 10,008 & 78 & 2,049.7 & 3,940.1 & 0.994 & 0.990 & 0.992 & 10.6 \\
\updated{FD7} & \updated{prior} & \updated{default} & \updated{\Y} & \updated{24} & \updated{1} & \updated{1} & \updated{10,065} & \updated{51} & \updated{1,708.1} & \updated{3,575.9} & \updated{0.989} & \updated{0.958} & \updated{0.973} & \updated{19.2} \\
FD8 & no-cert & default & \N & 27 & 0 & 2 & 9,991 & 21 & 1,518.3 & 3,467.4 & 0.995 & 0.982 & 0.988 & 12.8 \\
FD9 & no-cert & best & \Y & 27 & 0 & 2 & 10,003 & 26 & 1,518.3 & 3,503.6 & 0.996 & 0.990 & 0.993 & 13.6 \\

\bottomrule
\end{tabular}
\caption{Clustering results on the ground truth. ``Singl.'' refers to singleton clusters (i.e., only one element).}
\label{table:clustering_gt}
\end{table*}

\subsection{Clustering Results}
\label{sec:eval:clustering}

We leverage the ground truth to
assess the clustering accuracy along three dimensions:
features, 
clustering algorithm parameters, and 
z-score normalization for numerical features.
To evaluate clustering accuracy we use 
precision, recall, and F1 score, common metrics for evaluating 
malware clustering results~\cite{bayer2009scalable}.
These metrics do not require or use cluster labels. 
They measure structural similarity between the obtained clustering 
and the ground truth clustering. 

Table~\ref{table:clustering_gt} 
summarizes the results obtained for 9 clustering configurations.
First, we evaluate how different sets of features affect the results, 
using default parameters for the clustering and z-score normalization.
\updated{ 
We evaluate 7 sets of features.
FD1 corresponds to using all features and acts as a baseline. 
The other feature sets correspond to an ablation study where we remove 
some features and measure how much the accuracy metrics are reduced with 
respect to the FD1 baseline.
To evaluate the impact of each feature category, 
we build four feature sets (FD2--FD5), each excluding the features 
in one feature category
(e.g., excluding client features in FD2).
To evaluate the impact of payload features, 
we exclude features from the other three categories
(client, server, certificate) in FD6.
To evaluate the impact of our novel features, 
we exclude them in FD7, leaving only 
those already present in prior works.}
This search shows that server and payload features provide most information
(largest drop when excluded) and 
client features also provide some information.
However, certificate features are not useful 
(excluding them achieves best results) 
because they make the clustering split 
true clusters into subclusters due to the prevalence of 
certificate polymorphism and certificates from free CAs. 
The only useful information in free certificates is the domain name, 
which often overlaps the SNI feature. 
In addition, some CDNs like CloudFlare are abused by multiple families and 
the similarity of their certificates does not capture a real relationship.
Based on these results, when exploring the other
dimensions, we exclude the certificate features. 
\updated{The results from FD7 show that the new features proposed in 
this work increase the clustering accuracy raising the F1 score to 0.993, 
compared to 0.973 when using only the features proposed by prior work.}
Next, we evaluate the z-score normalization of the numerical features 
observing that normalization improves results.
Finally, we perform another search to optimize the clustering hyper-parameters.
The table does not show every parameter value configuration evaluated. 
It only shows the results with the best parameter configuration, 
which turns out to be the default parameters.
The best clustering is \updated{FD9}, 
which does not use certificate features,
applies z-score normalization for numerical features, and
uses hyper-parameters 
$\text{\it mpts}=2$, $\text{\it mcs}=2$, and $\text{\it m}=10$.
It achieves a precision of 0.996, a recall of 0.990, and
a F1 score of 0.993.

\begin{table}[t]
\centering
\footnotesize
\begin{tabular}{rrrrrrrr}
\toprule
\textbf{All} & \textbf{Singl.} & \textbf{Min} & \textbf{Max} & \textbf{Med.} & \textbf{Mean} & \textbf{PStdev} & \textbf{Time} \\
\midrule
	18,569 & 9,548 & 1 & 335,026 & 1 & 57.2 & 2,644.2 & 4.3h \\
\bottomrule
\end{tabular}
\caption{SB-all clusters using the best clustering (\updated{FD9}).}
\label{table:clustering}
\end{table}

\paragraph{SB-All clusters.}
Table~\ref{table:clustering} shows the clustering results on the 1M flows in
SB-all using the best clustering (\updated{FD9}).
It produces 18,569 clusters of which 
49\% contain multiple flows and the rest are singletons.
36\% of the clusters contain flows from multiple samples,
12\% more than one server leaf certificate,
and 3\% more than one SNI.
These results show that the clustering is able to group multiple 
samples that belong to the same family in the same cluster,  
despite domain and certificate polymorphism that malware
authors may apply to avoid blacklists.
We also observe that 31\% (5,847) of the clusters
(2,076 non-singleton clusters) only contain samples without 
an AVClass family label.
Samples in those families could not be detected using 
supervised multi-class classifiers since no labels exist for training 
their model.

We manually analyze the clusters and report below
some observations and example clusters.
The largest cluster has 335,026 flows from the \textit{clipbanker} family.
There are 1,118 clusters without domain information, 
i.e., all flows lack an SNI header. 
One such cluster has 8 flows, each from a different sample. 
The flows connect to three servers (i.e., destination IP addresses) 
that use two certificates, 
one for \textit{*.zohoassist.com} and the other for \textit{*.zoho.com}.
There are 8 content sequences, all with RRPs with similar sizes.
AVClass fails to provide a family label for all eight samples. 
This is an example of the clustering being able to group multiple samples
despite the absence of domain information and the malware using 
multiple certificates. 
It is also an example of a cluster where AVClass fails to label samples
and supervised multi-class classifiers do not work.

A total of 50 clusters (33 non-singletons) have TLS 1.3 flows, 
which lack certificates, and may also lack a SNI header, 
making them challenging to detect and label.
Of the 33 non-singleton clusters, 30 contain only TLS 1.3 flows and 
three contain multiple TLS versions.
Further examination shows that these 33 clusters likely belong to two 
families. 
According to AVClass, 
four TLS 1.3 clusters correspond to the \textit{sofacy} family. 
Each sample of this family produces a single TLS flow
to domains \textit{huikin.host} or \textit{w.huikin.host}, 
both hosted at IP address 18.197.147.148, 
but the clustering splits the traffic into subclusters with 
different encrypted payload sequences.
Of the 3,418 samples in these clusters, 37 have no AVClass family, 
but still are correctly clustered with their family peers.
For the other 29 non-singleton TLS 1.3 clusters,
we observe that all flows lack a SNI header, 
but go to the same destination IP address 3.123.117.231. 
Note that the destination IP address is not a feature in our clustering. 
This indicates that the grouping is correct although multiple clusters 
are obtained based on differences on the encrypted payload.
In the future, we plan to evaluate the destination IP address as a 
feature, which we originally thought was problematic due to IP reuse.
Many samples in these clusters have an AVClass family label, 
but those labels correspond to 17 families, 
so it is not possible to identify the correct family.
This highlights how our clustering can group samples not only with missing 
labels, but also with conflicting ones, 
which would be problematic for supervised approaches.

We observe multiple families abusing free certificates such as 
Let's Encrypt and Tencent Cloud's TrustAsia certificates.
One such cluster consists of 12 flows, 
each from a different sample. 
It contains three domains:
\textit{atendimentostore-al.com}, 
\textit{atendimentostore-ac.com}, and
\textit{buricamiudos-al.com}. 
The similarity between the domain names indicates that the cluster is correct.
Each domain resolves to a different server and 
has its own Let's Encrypt certificate. 
All samples have generic detection labels, 
so AVClass does not output a family for any of them.
The payload consists of four content sequences, 
all of them with two RRPs.
The first RRP has a 160 bytes request and a 7088--7280 bytes response. 
The second has a request 96--112 bytes request and 
a 5.18MB--5.38MB response, potentially corresponding to a downloaded file.
The clustering  is able to group the 12 samples 
despite the use of different domains and certificates. 

We also observe CloudFlare being abused by many families.
One such cluster consists of five flows from three samples. 
Each flow is for a different domain and goes to a 
different CloudFlare server using different CloudFlare-issued certificates.
The similarity in this case comes from the client and payload features. 
Overall, we observe that the clustering is able to split different families 
abusing CloudFlare infrastructure into their own clusters.

\begin{table}[t]
\centering
\footnotesize
\begin{tabular}{llrrrr}
\toprule
	\bf Model & \bf Thres. sel. & \bf Thres. & \bf MCS & \bf FDR & \bf Alarms \\
\midrule
	\updated{FD9} & var   &    - &  - & 1.8\%   & 1,698 \\
	\updated{FD9} & var   &    - & 50 & 0.11\%  &   109 \\
	\updated{FD9} & fixed & 0.20 &  - & 0.4\%   &   389 \\
	\updated{FD9} & fixed & 0.10 &  - & 0.08\%  &    73 \\
	\updated{FD9} & fixed & 0.05 &  - & 0.002\% &     2 \\
\bottomrule
\end{tabular}
\caption{Comparison of \mlalgorithm detector configurations on 95K flows from one day of benign traffic.}
\label{tab:detection_unsupervised_1day}
\end{table}

\subsection{Detection Results}
\label{sec:eval:detection}

This section evaluates our \mlalgorithm detector.
First, we select the best configuration.
Then, we measure the false detection rate (FDR) 
of the selected configuration 
using benign traffic from the research network.
Finally, we evaluate false negatives (FNs).

\paragraph{\Mlalgorithm detector configuration.}
For selecting the best \mlalgorithm detector configuration, 
we examine which threshold selection method produces a lower FDR 
with the best clustering configuration (\updated{FD9}), 
and whether removing small clusters
with few flows can significantly improve the FDR.
Table~\ref{tab:detection_unsupervised_1day} captures the results of 
applying each detector configuration on 95K flows from one day of 
benign traffic.
We first evaluate the variable threshold selection method. 
We compare results when keeping and removing clusters with 
less than 50 flows. 
Removing them reduces the FDR from 1.8\% to 0.11\%, 
but at the cost of not being able to detect traffic that matches 
those clusters. 
Then, we evaluate the fixed threshold selection method using 
different threshold values.
We start with threshold 0.2 because the vast majority of false alarms occur at that or greater
distances in the variable method.
Then, we halve that threshold a couple of times to observe the effects.
The results show that the fixed threshold achieves the lowest FDR and 
that smaller thresholds make the detection stricter and thus 
reduce FPs.
In the limit, a threshold of zero would make the detector flag only flows 
with identical feature vectors to those in the cluster. 
So, we select 0.05 as the best threshold as it still detects small 
modifications, as shown later in the FN evaluation.

\paragraph{FDR.}
To determine the real FDR of the \mlalgorithm detector, 
we first apply the selected configuration (0.05 fixed threshold) 
on one week of traffic from the Benign02 dataset,
observing a total of 119 alarms, 
for a FDR of 0.031\%.
To evaluate degradation over time, 
we then apply the detector again on almost four months of benign traffic
(111 days corresponding to union of the Benign02 and Benign03 datasets
in Table~\ref{tab:datasets_benign}),
observing 708 alarms produced by 5 clusters,
corresponding to a measured FDR of 0.032\%.
Thus, the FDR remains stable even after several months.

\paragraph{False negative evaluation.}
We also perform an experiment to validate that the chosen configuration of 
the \mlalgorithm detector is not too strict and still detects variations of 
flows in the clusters. 
For this experiment, we apply the clustering, 
using the same configuration as \updated{FD9},
on 90\% randomly sampled flows of each cluster in the ground truth, 
reserving the other 10\% as testing data never seen by the model.
We build an \mlalgorithm detector using the previously selected 
configuration of a fixed 0.05 threshold.
We use the produced model to make predictions on the 10\% unseen flows.
In this scenario, true positives (TPs) are flows assigned to a cluster
by a closest neighbor that shares the same manually assigned cluster in the 
ground truth.
If a flow is not assigned to any cluster (no nearest neighbor) it
is a FN, as we know it is malicious.
When the nearest neighbor of the flow is in a different ground truth 
cluster than the evaluated flow, 
we consider it a FP since the output cluster is wrong.
We perform a 10-fold cross-validation, 
sampling different testing data each time.
Five runs have 100\% TPs, 
and there are a total of 12 FNs on all 10 runs, 
for a FN rate of 0.029\%.
No FPs are observed.
This experiment confirms that the \mlalgorithm detector is capable of 
detecting previously unseen flows with low FNs, 
and that the selected clustering configuration is not too strict.

\subsection{Comparison with Prior Work}
\label{sec:eval:comparison}

\updated{This section compares \tool with two state-of-the-art 
publicly available detection tools: 
the Joy~\cite{joy} binary supervised classifier and 
Kitsune~\cite{kitsune}, an auto-encoder (AE) based anomaly detector.
Both tools can detect malicious traffic, 
but do not classify the detected traffic into families. 
Thus, the comparison focuses on the detection goal.
}

\paragraph{Joy.}
We compare our \mlalgorithm detector with Joy~\cite{joy}, 
the publicly available implementation of the binary supervised detector from 
Blake and McGrew~\cite{anderson2016identifying,anderson2017machine}.
We focus the comparison on the malicious flow detection
since the implementation of the multi-class supervised classifier 
by the same authors~\cite{anderson2016deciphering} is not publicly available.
One limitation of Joy is that it considers all network flows 
in the input network traces as malicious. 
However, as shown in Section~\ref{sec:datasets}, 
much traffic in sandbox traces is benign.
To understand the impact of this choice by Joy, 
as well as to make a fair comparison between Joy and our \mlalgorithm detector,
we build three different Joy logistic regression models.
The \emph{joy-polluted} model applies Joy without any modifications.
Thus, all flows in the sandbox network traces are considered malicious. 
The \emph{joy-unpolluted-exc} model excludes from the training 
benign flows identified by our filtering step.
The \emph{joy-unpolluted-inc} model labels the benign flows 
identified by our filtering step as negative. 
This last model explicitly tells Joy that the sandbox traces indeed 
contain benign traffic.
Finally, \emph{\mlalgorithm} corresponds to our 
\mlalgorithm detector using filtering and the best configuration
identified in Section~\ref{sec:eval:detection}.

For the training, we use 90\% of the network traces from the 
SB-small dataset as positive class 
(minus benign flows for models with filtering) 
and three days of traffic from the research network as the negative class. 
For the research network traffic, 
we run Joy in parallel with the TLS log collection since Joy requires 
network traces as input and cannot process directly our TLS logs.
For the testing, we use the remaining 10\% traces of SB-small and an 
extra day of traffic from the research network.
Two small differences in feature extraction between Joy and our approach are 
that Joy is not able to process a portion of SSLv2 flows and that it 
classifies any flow, even those that do not complete the TLS handshake.
To make the comparison as fair as possible, 
we exclude both cases so that the comparison happens on the same flows.

\begin{table}[t]
\centering
\footnotesize
\begin{tabular}{lrrrrrrr}
\toprule
{\bf Model} & \multicolumn{1}{c}{\bf TP} & \multicolumn{1}{c}{\bf FP} & \multicolumn{1}{c}{\bf TN} & \multicolumn{1}{c}{\bf FN} & \multicolumn{1}{c}{\bf Prec.} & \multicolumn{1}{c}{\bf Recall} & \multicolumn{1}{c}{\bf F1} \\
\cmidrule(r){1-1} \cmidrule(l){2-8}
	joy-polluted       & 19,950 & 142,703 & 284,836 &    11 & 0.12 & 0.99 & 0.22 \\
	joy-unpolluted-exc & 19,951 & 142,163 & 285,376 &    10 & 0.12 & 0.99 & 0.22 \\
	joy-unpolluted-inc & 15,515 &   2,154 & 425,385 & 4,446 & 0.88 & 0.77 & 0.82 \\
	\mlalgorithm       & 18,359 &   2,241 & 425,294 & 1,601 & 0.89 & 0.92 & 0.91 \\
\bottomrule
\end{tabular}
\caption{Comparison of Joy models with our approach.}
\label{tab:comparison}
\end{table}

Table~\ref{tab:comparison} summarizes the comparison.
Using Joy without modifications 
(joy-polluted) produces a very large 
number of FPs and overall a low 0.22 F1-score.
This indicates that Joy is learning to differentiate 
sandbox traffic (mostly TLS 1.0) 
from traffic from the research network (mostly TLS 1.2 and 1.3). 
This is confirmed by the difference between 
joy-unpolluted-exc and joy-unpolluted-inc. 
When, during training, Joy is told that benign traffic in the sandbox is 
not necessarily malicious (joy-unpolluted-inc), 
the accuracy greatly improves as many FPs are removed 
pushing the F1 score from 0.22 up to 0.82. 
However, FNs increase significantly because 
Joy is forced to relax its model that now has to account for 
some sandbox traffic not being malicious. 
In contrast, our \mlalgorithm approach 
captures different types of traffic from a family in their own clusters, 
producing a tighter model.
Our \mlalgorithm detector outperforms all Joy models, 
achieving a F1 score of 0.91, 
compared to 0.82 for Joy's best model. 

\begin{table}[t]
\centering
\footnotesize
\begin{tabular}{lrrrrrrr}
\toprule
{\bf Model} & \multicolumn{1}{c}{\bf TP} & \multicolumn{1}{c}{\bf FP} & \multicolumn{1}{c}{\bf TN} & \multicolumn{1}{c}{\bf FN} & \multicolumn{1}{c}{\bf Prec.} & \multicolumn{1}{c}{\bf Recall} & \multicolumn{1}{c}{\bf F1} \\
\cmidrule(r){1-1} \cmidrule(l){2-8}

	kitsune	(flow-level) &  15,681 & 13,137 &  3,527 &   8,323 & 0.54 & 0.65 & 0.59 \\
	\mlalgorithm	     &  30,297 &      1 & 22,317 &      94 & 0.99 & 0.99 & 0.99 \\
\bottomrule
\end{tabular}
\caption{\updated{Comparison with Kitsune on ground truth TLS traffic.}}
\label{tab:kitsune}
\end{table}

\paragraph{Kitsune.}
\updated{
We also compare \tool with Kitsune~\cite{kitsune}, 
a state-of-the-art AE-based anomaly detector.
Kitsune's approach does not require malicious 
traffic during training.
It builds an ensemble of neural network AEs from input benign traffic and 
derives a detection threshold by examining the maximum root-mean-square error (RMSE)
observed in the input benign traffic.
Given test traffic, Kitsune computes the RMSE value. 
If the value is larger than the inferred threshold, 
then the traffic is considered anomalous and an intrusion is flagged.
Similar to Joy, Kitsune takes as input network traces in PCAP format 
and thus we cannot use our TLS logs from the research network
for this evaluation.
Unlike with Joy, we could not run Kitsune in parallel with our TLS log 
collection, since our collection had been discontinued by the time we performed 
this evaluation.
To evaluate both tools on the same inputs, 
we leverage the sandbox traces.
We first filter out all non-TLS traffic from the traces
by selecting the packets to/from port tcp/443,
and then separate benign traffic from malicious traffic using our filtering 
(Section~\ref{sec:approach-filtering}).

At first, we tried to produce a Kitsune model using all the 
benign traffic in the SB-small dataset,
but after one week running, the model had not finished and we stopped it. 
Kitsune was designed to run in low-resource network devices 
and its design minimizes the amount of memory used.
However, it runs on a single thread making runtime the bottleneck. 
Thus, its model training does not scale to large amounts of traffic.
Because of this, we use a small set of ground truth traces
to train and test both tools.
More concretely, we use 4,125 ground truth traces for training each model.
Then, we test both tools on the same traffic that 
was not part of the training of neither of the two approaches. 
In particular, we use the benign traffic along with the malicious traffic of 5,156 traces
to build the testing dataset.
We use the Kitsune configuration suggested by its authors~\cite{kitsune}:
the detection threshold is the maximum 
RMSE value seen during the training phase
and the maximum number of features allowed per AE is 10.

Another key difference is that Kitsune operates at the 
packet level, while \tool operates at the flow level.
To compare the results of both approaches, 
we first create a mapping from each packet to the flow it belongs. 
When Kitsune flags a packet as an intrusion, 
we use the mapping to identify the flow to which the packet belongs 
and mark the whole flow as malicious.
Flows where no packet has been flagged as an intrusion are considered benign.
To build the packet-to-flow mapping, 
we produce a flow identifier for each packet using a set of six values: 
hash of the trace where the packet appears, 
source IP, 
source port, 
destination IP, 
destination port, and 
protocol.
To assign the same identifier to both directions of the flow 
we first sort the six values lexicographically and then hash them to 
produce the flow identifier.
After applying Kitsune's feature extractor to the ground truth traces, 
we add two additional fields to the Kitsune vectors:
the flow identifier and the SNI field for the flow. 
The SNI is used by the filtering to determine if a flow is benign or malicious.
We also modify the feature extraction of \tool 
to include the flow identifier.

Table~\ref{tab:kitsune} shows the results of both models.
Kitsune achieved a precision of 0.54,
a recall of 0.65,
and an F1 of 0.59.
On the same dataset, \tool achieves a precision of 0.99 and a recall 
of 0.99, for an F1 of 0.99.
The low precision of Kitsune is due to a large number of FPs (13,137). 
False positives are a common problem with anomaly detectors since 
any significant deviation from the profiled benign traffic is flagged as 
an intrusion, while the benign testing traffic may contain natural changes 
that are not related to malicious behavior.
Kitsune also introduces 8,323 FNs, 
likely due to malicious traffic with similar packet sizes and inter-arrival 
times to benign traffic, which cannot be easily distinguished 
using the benign traffic model.
\Tool can ameliorate that problem by using three additional 
categories of TLS-specific features (client, server, certificate).
}

\begin{table}[t]
\centering
\footnotesize
\begin{tabular}{|lc|cc|cccc|ccccccc|}
\cline{3-15}
\multicolumn{2}{c|}{} &
\multicolumn{2}{c|}{{\bf Goal}} & 
\multicolumn{4}{c|}{{\bf Approach}} &
\multicolumn{7}{c|}{{\bf Features}} \\
\hline
 \textbf{Work}
 &\textbf{Year}
 &\rcolumn{Detection}
 &\rcolumn{Classification}
 &\rcolumn{Supervised}
 &\rcolumn{Unsupervised}
 &\rcolumn{Anomaly Detection}
 &\rcolumn{Granularity}
 &\rcolumn{TLS Client}
 &\rcolumn{TLS Server}
 &\rcolumn{TLS Certificate}
 &\rcolumn{Payload Size}
 &\rcolumn{Inter-Arrival}
 &\rcolumn{DNS}
 &\rcolumn{HTTP} \\
\hline
APM~\cite{anderson2016deciphering} & 2016 & \Y & \Y & \Y & \N & \N & F & \Y & \Y & \Y & \Y & \Y & \N & \N \\ AM16~\cite{anderson2016identifying} & 2016 & \Y & \N & \Y & \N & \N & F & \Y & \Y & \Y & \Y & \Y & \Y & \Y \\ AM17~\cite{anderson2017machine} & 2017 & \Y & \N & \Y & \N & \N & F & \Y & \N & \N &\Y & \Y & \N & \N \\ Kitsune~\cite{kitsune} & 2018 & \Y & \N & \N & \N & \Y & P & \N & \N & \N & \Y & \Y & \N & \N \\
H2ID~\cite{bovenzi2020hierarchical} & 2020 & \Y & \Y & \Y & \N & \Y & F & \N & \N & \N & \Y & \Y & \N & \N \\
\hline
This work & 2022 & \Y & \Y & \N & \Y & \N & F & \Y & \Y & \Y & \Y & \N & \N & \N \\
\hline
\end{tabular}
\caption{\updated{Related work on detection and classification of malicious TLS flows.}
}
\label{tbl:related}
\end{table}

\section{Related Work}
\label{sec:related}

\updated{Table~\ref{tbl:related} summarizes the most related work. 
It includes three works that specifically explore the detection and 
classification of TLS flows~\cite{anderson2016deciphering,anderson2016identifying,anderson2017machine} 
and two general anomaly-based intrusion detection systems that do not 
specifically target TLS flows, but can be applied to them as they do 
not require access to unencrypted payload 
data~\cite{kitsune,bovenzi2020hierarchical}.
The bottom row captures the approach presented in this paper.
The table characterizes each work according to its goal, approach, and 
features used. 
The goal can be detecting malicious flows, 
as well as classifying those malicious flows by originating family.
The approach captures whether the work uses supervised ML models, 
unsupervised clustering, and 
anomaly detection based on auto-encoders. 
It also captures the approach granularity, 
i.e., whether it works at the flow (F) or packet (P) level.
The features used can be TLS-specific (client, server, certificate), 
specific to other protocols such as DNS and HTTP, or
examine the payload and inter-arrival times of packets in a flow 
in a protocol-agnostic manner.
Next, we compare these 5 works with our approach.
}

\citet{anderson2016deciphering} first propose a binary 
(logistic regression) supervised detector using features from TLS flows.
In addition, they train a logistic regression multi-class classifier 
for 18 malware families, 
but fail to generate a classifier for 73\% of their input samples 
for which they cannot obtain a family label. 
We compare our \mlalgorithm detector with the public implementation of
their binary supervised classifier. 
Their implementation assumes all TLS flows in the input network 
traces are malicious leading to a very low F1 score. 
After fixing that issue, the F1 score raises to 0.82, 
compared to 0.91 for our \mlalgorithm detector on the same data.
Compared to their multi-class classifier, 
our \mlalgorithm detector does not require family labels and 
detects flows from samples without a family, 
while still assigning a family to the detected flows if available. 

In follow-up work, \citet{anderson2016identifying} 
add to their TLS detector features from DNS responses 
and HTTP flows.
In contrast, our approach operates solely on TLS flows and does not require 
plain-text protocols, increasing user privacy. 
Furthermore, our \mlalgorithm approach can assign detected flows 
to their family clusters, and handles samples that do not use HTTP or  
connect to their C\&C servers using IP addresses.

\citet{anderson2017machine} also evaluate
six supervised learning classifiers 
for detecting malicious TLS flows despite 
inaccurate ground truth and non-stationarity of network data.
In their experiments, 
random forest performed best, but the quality of its results decreased 
significantly over time.
In comparison, the FDR of our \mlalgorithm detector does not degrade 
over a four-month period.

\updated{
\citet{kitsune} present Kitsune, an anomaly detection approach that 
takes as input benign traffic and uses an ensemble of auto-encoders 
to detect anomalies that indicate intrusions.
Using an anomaly-based approach removes the need for a malicious training 
dataset, but prevents it from addressing the classification goal.
Kitsune uses packet sizes and inter-arrival times as features, 
so it can be applied to any protocol including TLS flows.
It is designed to be efficient, so that it can run on network devices that 
have limited resources (e.g., a Raspberry PI).
We compare our approach with the publicly available implementation of Kitsune
showing that our approach significantly improves the accuracy.

\citet{bovenzi2020hierarchical} present H2ID, an intrusion detection system 
that addresses both the detection and classification goals. 
Similar to Kitsune, it first uses an auto-encoder approach to detect 
anomalies as intrusions and uses payload size and inter-arrival timing 
as features, so it can be applied to different protocols, including TLS flows.
But, it adds a second phase where it applies a supervised ML model to 
classify the detected anomalies into known attacks, 
or unknown if they do not match the model.
In contrast, our approach does not require a labeled training dataset and 
can cluster similar malicious flows even when their family is unknown.
}

\paragraph{TLS fingerprints.}
A large body of work has built TLS fingerprints to identify the
applications that initiate TLS flows.
Fingerprints have been used to analyze several aspects of the TLS ecosystem, including 
the impact of HTTPS interception by middleboxes and antivirus products~\cite{durumeric2017security},
the evolution of TLS clients over time~\cite{kotzias2018coming, anderson2019tls},
and the TLS implementations of popular censorship circumvention tools~\cite{frolov2019use}.
Prior works also build TLS fingerprints for detecting malware and 
PUP families~\cite{leebrotherston,salesforcetls}.
However, recent work shows that malware TLS fingerprints 
generate high FPs in real networks~\cite{anderson2019tls}.
All these approaches build TLS fingerprints using solely features 
extracted from the Client Hello message such as TLS version,  
supported ciphersuites and extensions, and 
elliptic curves point formats.
Van Ede et al.~\cite{vanede2020flowprint} proposed FlowPrint, a semi-supervised
approach for fingerprinting mobile apps from their TLS traffic using also
destination features such as the server certificate, IP address, and port.
In comparison, our clustering also uses client and certificate features, 
but enhances the detection by further including server 
and encrypted payload features.

\paragraph{Malware clustering.}
There has been extensive work on malware clustering techniques 
using a variety of features such as system calls, system changes, and
network traffic~\cite{Bailey07,Rieck08,botminer,bayer2009scalable,Perdisci10,dietrich2013,firma,buyukkayhaa2017n}. 
Most similar to our approach are clustering approaches based 
on network traffic, 
which may build detection signatures~\cite{botminer,Perdisci10,firma} 
or unsupervised detectors~\cite{dietrich2013} using the produced clusters.
Some of these works propose generic payload features
that capture message sizes~\cite{dietrich2013,firma}, 
but none of these works uses TLS-specific features.
In future work, we would like to combine our TLS features with other 
network and behavioral features to build a malware family classifier 
for samples executed in a sandbox.

\section{Discussion}
\label{sec:discussion}

This section discusses limitations and avenues for improvement.

\paragraph{Potential biases.}
We identify two potential sources of bias in our methodology.
First, we discard almost 64\% of the sandbox TLS flows in our malware
dataset because they try to use TLS 1.0, 
which was already deprecated by many servers at the time of the execution.
This happens because most malware uses the Windows TLS functionality, and
Windows 7, the sandbox OS version, uses by default TLS 1.0.
Of the 9.3M discarded connections, 23\% would have been anyway removed
later by the filtering. Second, we filter out benign domains, which may result
in excluding some malware families from our study.
For example, click-fraud malware generates revenue by falsifying clicks 
on pay-per-click campaigns and typically contacts numerous benign destinations.
Other malware families may abuse benign content hosting services such as
code repositories (e.g., Github and Bitbucket), document editors (e.g.,
Google Docs, Pastebin), cloud storage services (e.g., Dropbox, Google Drive),
or even social media (e.g., Twitter).
Despite excluding some malware families, our dataset still encompasses a large
part of the malware ecosystem, containing 738 different families
(see Section~\ref{sec:classification}).
A third bias could be introduced by our ground truth not being representative 
of the whole SB-all dataset. 
We try to ameliorate this by including multiple families, 
limiting the number of samples of each family, and 
including samples without a known family.

\paragraph{Detection through other protocols.}
Usage of TLS by malware keeps increasing, having more than doubled 
from 10\% in 2016~\cite{anderson2016deciphering}
up to 23\% in 2020~\cite{mal_tls_adoption}.
However, malware may still use unencrypted traffic or 
mix unencrypted and encrypted protocols, 
as illustrated by 91\% of samples using plain HTTP in addition to TLS. 
Prior work suggested using unencrypted protocols such as 
HTTP and DNS to assist in the detection of 
TLS malware~\cite{anderson2016identifying}. 
However, we already observe 9\% of the samples using TLS, 
but not other unencrypted traffic beyond DNS; 
we expect this ratio to keep increasing over time.
Moreover, samples not using (or encrypting) a TLS SNI header 
(the one case where DNS helps) could also connect 
to their C\&C using IP addresses.
Furthermore, a purely TLS-based detection improves user privacy, and 
technologies such as 
DNS over HTTPS could further hamper DNS utility.
While detection via other protocols is still a possibility for a significant 
fraction of samples, 
that is not the case for all samples and the situation is 
likely to get worse.

\section{Conclusion}
\label{sec:conclusion}

Detecting malicious TLS traffic is an important, 
but challenging, problem.
Binary supervised detectors are limited in that they try to
distinguish any malicious TLS flow,
regardless of the malware family producing it.
This produces models that are too loose, introducing errors.
In addition, they do not provide contextual information about the 
detected flows.
Multi-class supervised classifiers produce tighter models for specific 
malware families, and can classify the detected flows. 
But, they require family labels to train the classifiers, 
which are not available for a large fraction of samples. 

We have proposed a novel \mlalgorithm approach for 
detecting and clustering malicious TLS flows.
It first clusters similar malicious TLS flows without requiring family labels.
Then, it builds an \mlalgorithm detector that 
measures distance to the clusters to determine 
if a given flow is malicious 
(belongs to a cluster)
or benign (no cluster is close enough).
We have evaluated our approach using 972K traces from a commercial sandbox 
and 35M TLS flows from a research network.
Our unsupervised detector achieves a F1 score of 0.91,
compared to 0.82 for the state-of-the-art supervised detector,
and a FDR of 0.032\% over four months of traffic.

\updated{
In future work, we plan to evaluate additional features such as 
the destination IP address and timing-related features.
For example, the IP address we originally thought was problematic 
due to IP reuse, 
but our evaluation shows there are cases where it can be useful. 
Timing-related features have been used by prior work 
(e.g.,~\cite{anderson2016deciphering,kitsune,bovenzi2020hierarchical}),
but our examination found them too sensitive to the specific network setup, 
so we believe there is a need for a systematic evaluation on their usefulness.
Furthermore, we would like to explore how to generalize \tool to handle other 
types of network traffic beyond TLS.
}

\section{Data Availability Statement}

This work uses two data types: 
[MALWARE-TRACES] are network traces from likely malicious samples executed 
in a sandbox. 
[BENIGN-TRACES] are network traces from a research network containing traffic considered benign.
MALWARE-TRACES data used to support the findings of this study have not been made available because they come from a commercial sandbox
and are considered proprietary data for the commercial service.
BENIGN-TRACES data used to support the findings of this study have not been made available because they contain sensitive user data that cannot be made 
publicly available by the research institution.

\section{Acknowledgments}

This research was funded by the Madrid regional government
program S2018/TCS-4339 (BLOQUES-CM) and
by the Spanish Governement MCIN/AEI/10.13039/501100011033/ERDF 
through the SCUM project (RTI2018-102043-B-I00) 
the PRODIGY project (TED2021-132464B-I00), 
and an FPI grant (PRE2019-088472).
Those projects are co-funded by
European Union ESF, EIE, and NextGeneration funds.

\updated{
This submission is also available as an Arxiv preprint~\cite{tr2021}.
}

\appendix
\section{Appendix}

\begin{table}[h]
\centering
\footnotesize
\begin{tabular}{llrrrr}
\toprule
\textbf{Family} & \textbf{Type} & \textbf{SNI} & \textbf{IP} & \textbf{TLS flows} & \textbf{Samples} \\
  \midrule
clipbanker        & Malware & 7   & 10  & 319,185 & 88,680 \\
upatre            & Malware & 251 & 347 & 311,687 & 64,860 \\
shiz              & Malware & 12  & 18  & 41,578  & 19,959 \\
zbot              & Malware & 120 & 116 & 46,437  & 5,502  \\
pcchist           & Malware & 1   & 1   & 4,713   & 4,653  \\
installmonster    & PUP     & 61  & 75  & 4,537   & 4,163  \\
sofacy            & Malware & 1   & 1   & 3,381   & 3,381  \\
xetapp            & PUP     & 2   & 2   & 3,908   & 2,828  \\
oxypumper         & PUP     & 7   & 502 & 3,051   & 2,431  \\
bublik            & Malware & 36  & 33  & 30,741  & 2,240  \\
vkontaktedj       & PUP     & 2   & 12  & 2,060   & 1,599  \\
khalesi           & PUP     & 8   & 7   & 1,787   & 1,591  \\
playtech          & PUP     & 8   & 17  & 2,912   & 1,497  \\
multiplug         & PUP     & 2   & 2   & 9,223   & 1,055  \\
adposhel          & PUP     & 2   & 28  & 958    & 958   \\
razy              & Malware & 88  & 120 & 10,579  & 945   \\
lockyc            & Malware & 2   & 3   & 724    & 709   \\
noobyprotect      & Malware & 12  & 22  & 1,133   & 632   \\
downloadassistant & PUP     & 50  & 63  & 574    & 573   \\
delf              & Malware & 76  & 114 & 1,127   & 542   \\
\bottomrule
 \end{tabular}
	\caption{Top 20 families by samples after filtering.}
\label{tab:top_fams}
\end{table}

\begin{table}
\centering
\footnotesize
\begin{tabular}{|ll|}
\hline
	\textbf{Acronym} & \textbf{Full name} \\
\hline
AE & Auto-Encoder \\
ALPN & Application Layer Protocol Negotation TLS extension \\
API & Application Programming Interface \\
APT28 & Fancy Bear Russian cyber espionage group \\
C\&C & Command-and-Control \\
CA & Certification Authority \\
CDN & Content Delivery Network \\
CT & Certificate Transparency \\
DHCP & Dynamic Host Configuration Protocol \\
DN & Distinguished Name \\
DNS & Domain Name System \\
e2LD & Effective Second-Level Domain \\
FDR & False Detection Ratio \\
FN & False Negative \\
FP & False Positive \\
HDBSCAN & Hierarchical Density-Based Spatial Clustering of Applications with Noise \\
HTTP & Hypertext Transfer Protocol \\
HTTPS & Hypertext Transfer Protocol Secure \\
IP & Internet Protocol \\
mcs & minimum cluster size \\
MITM & Man-in-the-Middle \\
ML & Machine learning \\
OS & Operating System \\
PUP & Potentially Unwanted Programs \\
RMSE & Root-mean-square Error \\
RRP & Request-Response Pair \\
SB & sandbox \\
SAN & Subject Alternative Name TLS extension \\
SNI & Server Name Indication TLS extension \\
SSL & Secure Sockets Layer \\
TF-IDF & Term-Frequency Inverse Document-Frequency \\
TCP & Transmission Control Protocol \\
TLS & Transport Layer Security protocol \\
TP & True Positive \\
URI & Uniform Resource Identifier \\
VT & VirusTotal \\
\hline
\end{tabular}
\caption{\updated{Acronyms used in the paper.}}
\label{tab:acronyms}
\end{table}

\bibliographystyle{ACM-Reference-Format}
\bibliography{bibliography/paper}

\end{document}